\documentclass[aps,floatfix]{revtex4}
\usepackage{graphics,epsfig}
\usepackage{graphicx}

\begin{document}

\title{A tool for teaching General Relativity}
\author{Kayll Lake \cite{email}}
\affiliation{Department of Physics and Department of Mathematics
and Statistics, Queen's University, Kingston, Ontario, Canada, K7L
3N6 }
\date{\today}

\begin{abstract}
GRTensorJ - Books is an active interface to a small part of the
computer algebra systems GRTensorII (for Maple) and GRTensorM (for
Mathematica) with the specific intent of providing students of
General Relativity with an advanced programmable calculator-style
tool. All standard functions associated with a classical tensor
approach to the subject are available thus reducing these to
``elementary functions". This is not a traditional database. The
database entries are spacetimes and calculations are done in real
time. All spacetimes are referenced directly by equation number in
ten current (and classic) texts in notation as close as possible
to the original text. The tool is now available free of charge
from \texttt{grtensor.org/teaching} .
\end{abstract}
\maketitle

\section{Motivation}

Recently Hartle \cite{hartle} has emphasized the increasing
importance of General Relativity on the frontiers of both the very
largest and the very smallest distance scales, a fact reflected by
the increasing number of undergraduate physics majors being
introduced to the subject. Of course, one can only learn physics
by doing it, but many of the standard functions of General
Relativity, when taught through classical tensor methods, involve
only partial differentiation and summation. I believe that little
is to be gained by repetitive routine calculations of this nature.
For this reason, I would like to introduce here a tool that will
be of use for all students starting their study of General
Relativity. The purpose of the tool is to turn all standard
(tensor based) functions into ``elementary functions" - those
immediately available with the click of a button - so that the
inner beauty of General Relativity shines through.

\section{Background}

The tool described here involves and update \cite{update} to the
Java \cite{java} interface GRTensorJ \cite{grtj}. The workings of
the interface have been described in detail previously
\cite{ishak}. The version of GRTensorJ described here cannot be
run over the internet but rather is a local application. The
interface currently covers every spacetime (referenced by equation
number and in notation as close to the original as possible) in
the following texts:

\bigskip

 \begin{itemize}
 \item Carroll, Sean. \textit{Spacetime and Geometry: An Introduction to General
 Relativity}, Addison Wesley (San Francisco), 2004. ISBN
 0-8053-8732-3
 \item d'Inverno, Ray. \textit{Introducing Einstein's Relativity}, Clarendon Press (Oxford), 1992. ISBN
 0-19-859686-3
 \item Hartle, James. \textit{Gravity: An Introduction to Einstein's General Relativity} , Addison Wesley (San Francisco), 2003.
    ISBN 0-8053-8662-9
 \item Hawking, Stephen, and G.F.R. Ellis. \textit{The Large Scale Structure of
 Spacetime }, Cambridge University Press (Cambridge), 1973. ISBN
 0-521-09906-4
 \item Misner, Charles, Kip Thorne and John Wheeler. \textit{Gravitation} , W.H. Freeman and Co. (San Francisco), 1973.
    ISBN 0-7167-0344-0
 \item Poisson, Eric. \textit{A Relativist's Toolkit: The Mathematics of Black-Hole Mechanics} , Cambridge University
    Press (Cambridge), 2004. ISBN 0-521-83091-5
 \item Rindler, Wolfgang. \textit{Relativity: Special, General and Cosmological} , Oxford University Press (Oxford), 2001.
    ISBN 0-19-850836-0
 \item Schutz, Bernard. \textit{A First Course in General Relativity} , Cambridge University Press (Cambridge), 1990.
    ISBN 0-521-27703-5
 \item Wald, Robert. \textit{General Relativity }, University of Chicago Press (Chicago), 1984. ISBN
 0-226-87033-2
 \item Weinberg, Steven. \textit{Gravitation and Cosmology: Principles and Applications of the General Theory of
 Relativity}, John Wiley and Sons (New York), 1972. ISBN
 0-471-92567-5
 \end{itemize}

\section{Preprogrammed Calculations}

In addition to a complete help system, the initial set of preprogrammed objects available for coordinate
calculations is shown in TABLE 1. Note that the table is condensed
as, for example, all index configurations are immediately
available.  Note also that context - sensitive mathematical
information is embedded as explained below.  The introduction of
further objects is also described below.

\begin{table}
\begin{tabular}{lll}

\textbf{GRTensorJ menu item}  & \textbf{GRTensorJ sub-menu
command} & \textbf{Object(s)}\\ \\

Metric  & Metric                       & $g_{ab}$\\

        & Signature                    & $Sig$\\

        & Line Element                 & $ds^{2}$\\

        & Constraints                             \\

        & Determinant of Metric        & $det(g)$\\

        & Inverse Metric               & $g^{ab}$\\

        & Partial Derivative of Metric & $g_{ab,c}$\\

Christoffel Symbols & Chr(dn,dn,dn)    & $\Gamma_{abc}$\\

                    & Chr(dn,dn,up)    & $\Gamma^{c}_{ab}$\\

Geodesic            &      &Writes out the geodesic equations      \\

Riemann             & R(dn,dn,dn,dn)   & $R_{abcd}$\\

                    & R(up,dn,dn,dn)   & $R^a\,_{bcd}$\\

                    & R(up,up,dn,dn)   & $R^{ab}\,_{cd}$\\

                    & R(up,up,up,up)   & $R^{abcd}$\\

                    & Kretschmann      & $R_{abcd}R^{abcd}$\\

Weyl                & C(dn,dn,dn,dn) ... & $C_{abcd}$\\

                    & Dual CStar(dn,dn,dn,dn)

... & $C^{*}_{abcd}\equiv\frac{1}{2}\epsilon_{abef}C^{ef}\,_{cd}$ \\

Ricci               & R(dn,dn) ...     & $R_{ab}\equiv R^c\,_{acb}$\\

Trace-free Ricci    & S(dn,dn) ...     & $S_{ab}\equiv R_{ab}-\frac{1}{N}g_{ab}R$\\

Einstein            & G(dn,dn) ...     & $G_{ab}\equiv R_{ab}-\frac{1}{2}g_{ab}R$\\

Invariants          & Invariants-Ricci & $R\equiv R^{a}\,_{a},\, R1\equiv\frac{1}{4}S^{a}\,_{b}S^{b}\,_{a}, \; R2\equiv\frac{-1}{8}S^{a}\,_{b}S^{b}\,_{c}S^{c}\,_{a}$ \\

                    &                  & $R3\equiv\frac{1}{16}S^{a}\,_{b}S^{b}\,_{c}S^{c}\,_{d}S^{d}\,_{a}$ \\

                    & Invariants-Weyl  & $W1R\equiv\frac{1}{8}C_{abcd}C^{abcd}, \; W1I\equiv\frac{1}{8}C^{*}\,_{abcd}C^{abcd}$ \\

                    &                  & $W2R\equiv\frac{-1}{16}C_{ab}\,^{cd}C_{cd}\,^{ef}C_{ef}\,^{ab}$ \\

                    &                  & $W2I\equiv\frac{-1}{16}C^{*}\,_{ab}\,^{cd}C_{cd}\,^{ef}C_{ef}\,^{ab}$ \\

                & Invariants-Mixed & $M1R \equiv \frac{1}{8}S^{ab}S^{cd}C_{abcd},\, M1I\equiv

\frac{1}{8}S^{ab}S^{cd}C^{*}\,_{abcd}$ \\

                    &                  & $M2R \equiv \frac{1}{16}S^{cd}S_{ef}(C_{acdb}C^{aefb}-C^{*}_{acdb}C_{*}^{aefb})$ \\

                    &                  & $M2I \equiv \frac{1}{8}S^{bc}S_{ef}(C^{*}_{abcd}C^{aefd})$ \\

                    &                  & $M3 \equiv \frac{1}{16}S^{cd}S_{ef}(C_{acdb}C^{aefb}+C^{*}_{acdb}C_{*}^{aefb})$ \\

                    &                  & $M4 \equiv \frac{-1}{32}S^{cg}S^{ef}S^{c}\,_{d}(C_{ac}\,^{db}C_{befg}+C^{*}_{ac}\,^{db}C^{*}_{befg})$ \\

              &                  & $M5R \equiv \frac{1}{32}S^{cd}S^{ef}C^{aghb}(C_{acdb}C_{gefh}+C^{*}_{acdb}C^{*}_{gefh})$\\

                    &                  & $M5I \equiv \frac{1}{32}S^{cd}S^{ef}C_{*}\,^{aghb}(C_{acdb}C_{gefh}+C^{*}_{acdb}C^{*}_{gefh})$\\

                    &                  & $M6R \equiv \frac{1}{32} S_a{}^e S_e{}^c S_b{}^f

                  S_f{}^d C^{ab}{}_{cd}$  \\

                    &                 & $ M6I \equiv \frac{1}{32} S_a{}^e S_e{}^c S_b{}^f

                  S_f{}^d C^{*ab}{}_{cd}$  \\

Differential Invariants& diRicci         & $R_{ab;c}R^{ab;c}$\\

                       & diRiem          & $R_{abcd;e}R^{abcd;e}$\\

                       & diS             & $S_{ab;c}S^{ab;c}$\\

                       & diWeyl          & $C_{abcd;e}C^{abcd;e}$\\

Bel-Robinson           & T(dn,dn,dn,dn) ... &

$T_{cdef}\equiv C_{acdb}C^{a}\,_{ef}\,^{b}+C^{*}\,_{acdb}C^{*}\,^{a}\,_{ef}\,^{b}$ \\

Weyl-Schouten          & Weyl-Schouten ...  &

$WS_{abc}\equiv R_{ab;c}-R_{ac;b}-( g_{ab}R^{e}\,_{e;c}-g_{ac}R^{f}\,_{f;b} )/4$ \\

Bach                   & B(dn,dn) ...       & $B_{ac}\equiv

C_{abcd}\,^{;bd}+\frac{1}{2}R^{bd}C_{abcd}$\\

\end{tabular}
\caption{Initial set of preprogrammed objects available for
coordinate calculations (condensed). Note: ``..." means that all
index combinations are immediately available as with Riemann.}
\end{table}

\section{Loading a Spacetime and Working with it}

\begin{figure}[ht]
\epsfig{file=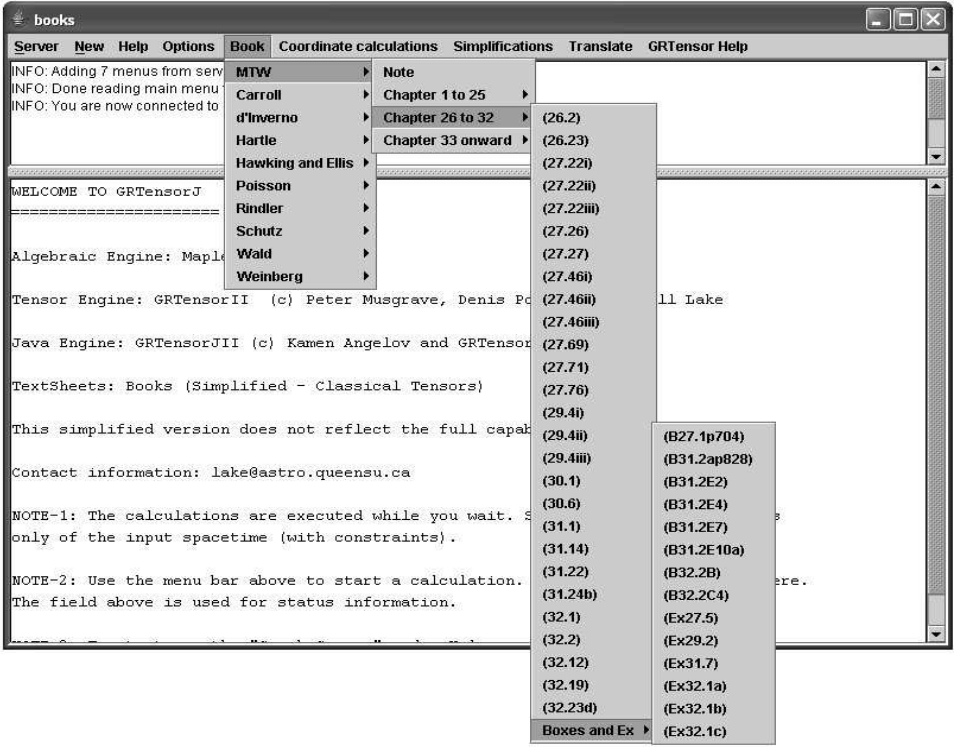,height=4in,width=5in,angle=0}
\caption{\label{book2a} A look at the middle section of the
classic text by Misner, Thorne and Wheeler.}
\end{figure}

FIG.~\ref{book2a} shows a screen shot of the interface running
Maple 10 \cite{maple} under Windows XP \cite{micro} and samples
the coverage associated with the classic text by Misner, Thorne
and Wheeler. Assuming you have the text, you simply click on the
equation number to load the spacetime.

\bigskip

\begin{figure}[ht]
\epsfig{file=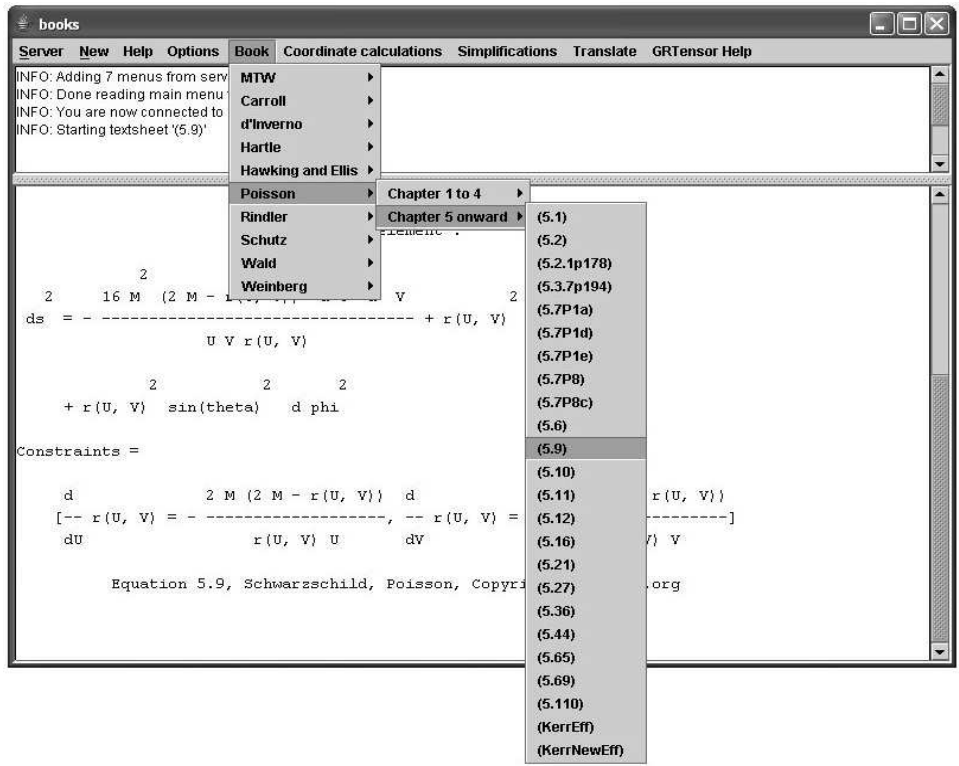,height=4in,width=5in,angle=0}
\caption{\label{book1} Loading of the Kruskal - Szekeres covering
of the Schwarzschild vacuum as discussed by Poisson.}
\end{figure}

\begin{figure}[ht]
\epsfig{file=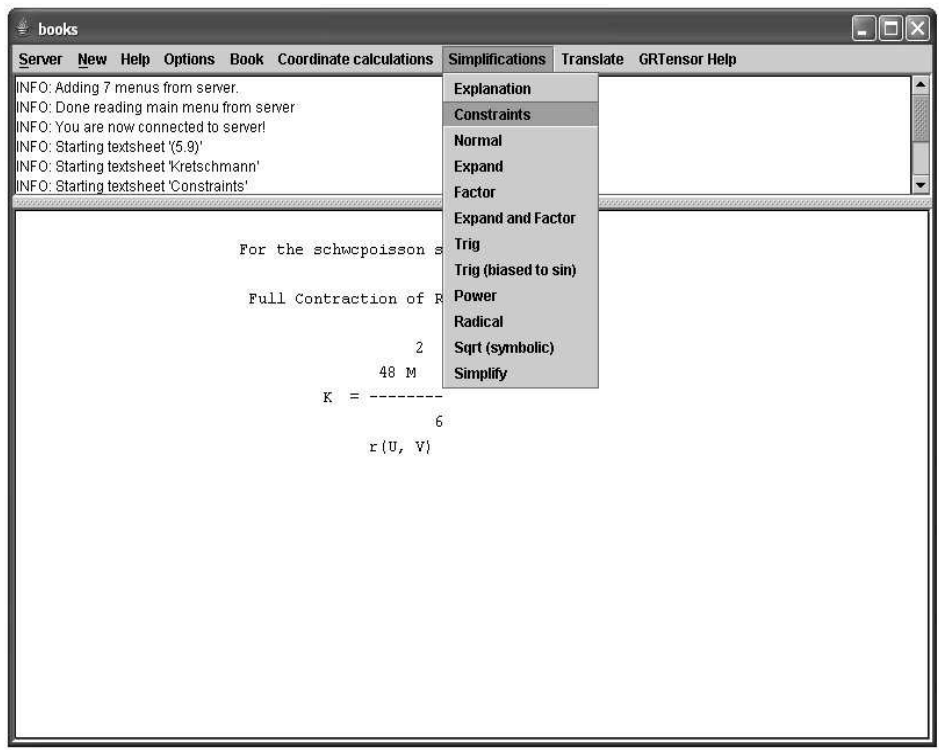,height=4in,width=5in,angle=0}
\caption{\label{book2} Applying constraints to the calculation of
the Kretschmann scalar (see Table 1).}
\end{figure}

FIG.~\ref{book1} shows loading of the Kruskal - Szekeres covering
of the Schwarzschild vacuum as discussed by Poisson. We use this
example here because the coordinates are implicitly defined and
therefore involve constraints which can be seen in the figure.
Without these constraints the spacetime is not defined.
FIG.~\ref{book2} shows the result of applying constraints to the
calculation of the Kretschmann scalar (see TABLE 1).

\bigskip

\begin{figure}[ht]
\epsfig{file=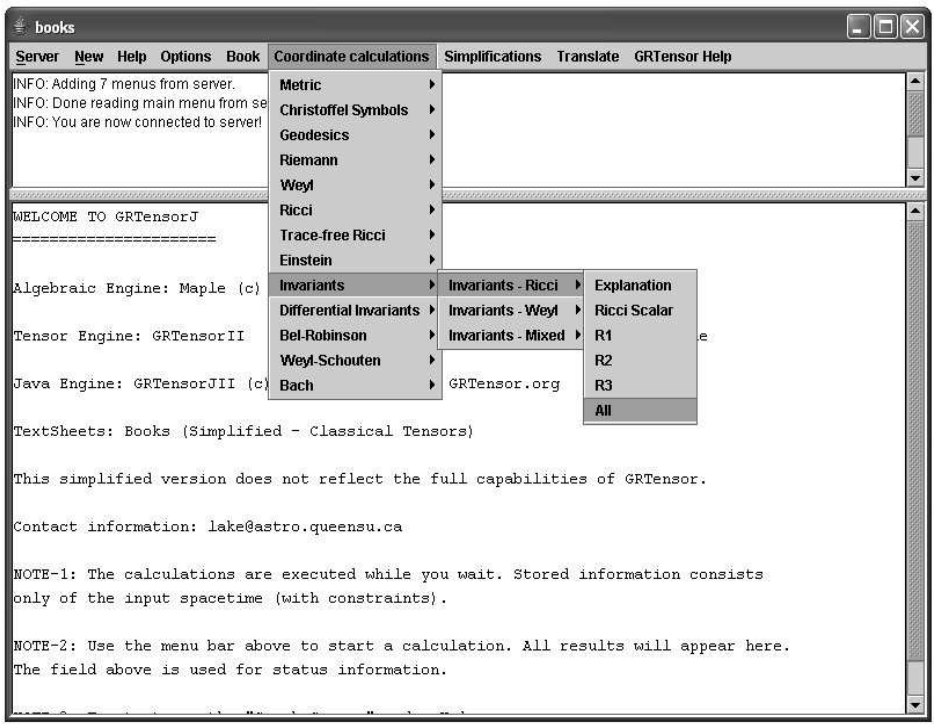,height=4in,width=5in,angle=0}
\caption{\label{book3} Calculation of Ricci invariants.}
\end{figure}

\begin{figure}[ht]
\epsfig{file=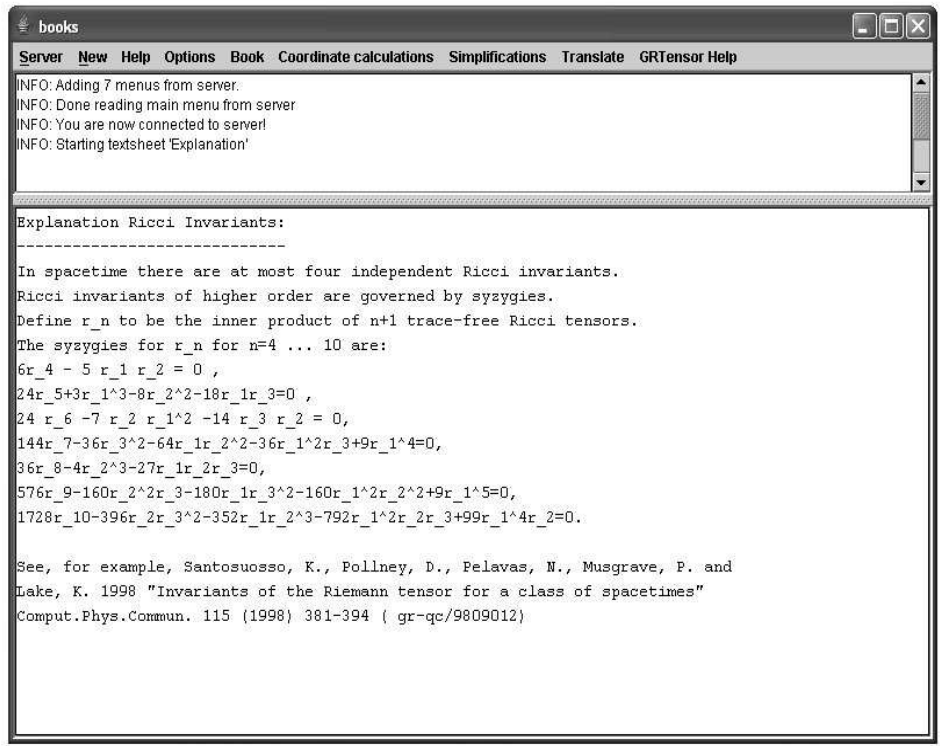,height=4in,width=5in,angle=0}
\caption{\label{book4} Explanation for the choice of Ricci
invariants. }
\end{figure}

As mentioned above, the interface contains context - sensitive
embeded mathematical information. An example is shown in
FIG.~\ref{book3} where one notices that there are only four Ricci
invariants available for calculation. The reason for this is given
in the explanation as shown in FIG.~\ref{book4} \cite{ricci}.

\section{Interface Expansion}

At the time of writing, virtually any calculation one encounters
in classical tensor analysis is already programmed into the
interface. Further routines can be easily added through the
GRTensorII \texttt{grdef} and GRTensorM \texttt{grdefine}
facilities. This includes tetrad calculations which we do not
discuss here. To see how to program the interface, let us look at
the construction of the Weyl-Schouten tensor (equivalently, the
Cotton-York tensor which is also available). This is a three index
tensor which in three dimensions is zero if and only if the space
is conformally flat. As can be see from TABLE 1, it is defined by
\begin{eqnarray}
WS_{abc}\equiv R_{ab;c}-R_{ac;b}-(
g_{ab}R^{e}\,_{e;c}-g_{ac}R^{f}\,_{f;b} )/4, \nonumber
\end{eqnarray}
where the terms are defined in the table. This relation is
programmed into the interface by including a plain text file
(Weyl-Scouten.ts) the entire contents of which are:

 \begin{verbatim}

grdef(`WS{a [ b c ] }:=R{a b ; c}-R{a c ; b}-(g{a b}*R{^e e;c}-g{a c}*R{^f f ; b})/4`);

grcalc(WS(dn,dn,dn));

*grdisplay(_);

 \end{verbatim}

\section{System Requirements}

To run GRTensorJ - Books you need Java \cite{java} installed on
your computer and a version of Maple or Mathematica. (You do not
need to have GRTensorII or GRTensorM already installed.)  The
Maple version (tested on Maple 7, 8, 9, 9.5.2, 10 and student 10)
for Windows (tested on 2000 and XP) is available now from
\texttt{grtensor.org/teaching}. Versions for other platforms, and
updates, will be announced at the web site as they are available.

\bigskip

\begin{acknowledgments}
James Atkinson, Alex Johnstone and Reuble Mathew helped with the
metric files but all errors are my responsibility. This work was
supported in part by a grant from the Natural Sciences and
Engineering Research Council of Canada.
\end{acknowledgments}

\bigskip

\bigskip

\begin{appendix}

\end{appendix}

\begin{thebibliography}{}
\bibitem{update} The update to the interface was written by Kamen
Angelov.
\end{thebibliography}
\end{document}